\documentclass[aps,pra,reprint,groupedaddress]{revtex4-1}
\usepackage[UKenglish]{babel}
\usepackage[]{graphicx, amsfonts, amsmath, amssymb, amstext, latexsym, float, color}

\newcommand{\be}{\begin{equation}}
\newcommand{\ee}{\end{equation}}
\newcommand{\ber}{\begin{eqnarray}}
\newcommand{\eer}{\end{eqnarray}}
\newcommand{\Tr}{\text{Tr}}

\newcommand{\ket}[1]{|#1\rangle}	
\newcommand{\bra}[1]{\langle#1|}
\newcommand{\braket}[2]{\langle#1|#2\rangle}

\newcommand{\mx}[1]{\begin{pmatrix}#1\end{pmatrix}}

\begin{document}

\title{Optimal Measurements for Symmetric Quantum States with Applications to Optical Communication}

\author{Hari Krovi, Saikat Guha, Zachary Dutton, Marcus P. da Silva}
\affiliation{Quantum Information Processing group, Raytheon BBN Technologies, 10 Moulton Street, Cambridge, MA USA 02138}


\begin{abstract}
The minimum probability of error (MPE) measurement discriminates between a set of candidate quantum states with the minimum average error probability allowed by quantum mechanics. Conditions for a measurement to be MPE were derived by Yuen, Kennedy and Lax (YKL)~\cite{YKL}. MPE measurements have been found for states that form a single orbit under a group action, i.e., there is a transitive group action on the states in the set. For such state sets, termed {\em geometrically uniform} (GU) in~\cite{Forney}, it was shown that the `pretty good measurement' (PGM) attains the MPE. Even so, evaluating the actual probability of error (and other performance metrics) attained by the PGM on a GU set involves inverting large matrices, and is not easy in general. Our first contribution is a formula for the MPE and conditional probabilities of GU sets, using group representation theory. Next, we consider sets of pure states that have multiple orbits under the group action. Such states are termed {\em compound geometrically uniform} (CGU). MPE measurements for general CGU sets are not known. In this paper, we show how our representation-theoretic description of optimal measurements for GU sets naturally generalizes to the CGU case. We show how to compute the MPE measurement for CGU sets by reducing the problem to solving a few simultaneous equations. The number of equations depends on the sizes of the multiplicity space of irreducible representations. For many common group representations (such as those of several practical good linear codes), this is {\em much} more tractable than solving large semi-definite programs---which is what is needed to solve the YKL conditions numerically for arbitrary state sets. We show how to evaluate MPE measurements for CGU states for some examples relevant to quantum-limited classical optical communication.
\end{abstract}

\maketitle

\section{Introduction}
Optimal discrimination of quantum states is central to a large number of key problems in quantum information theory. Quantum state discrimination finds applications in: (i) {\em computation}---for instance, in quantum algorithms for hidden subgroup problems~\cite{Bac05, Moore_Russell}; (ii) {\em sensing}---for instance, in task-specific optical imaging~\cite{Tan08}, quantum reading~\cite{Pir11}, and pixelated image discrimination~\cite{Nai11}; (iii) {\em communication}---for instance, in decoding error correcting codes for classical communication over a quantum optical channel~\cite{Llo10, Guh11, Wil12a, Wil12} and optimal $M$-ary phase discrimination under a photon budget constraint~\cite{Nai12}. The problem of describing a quantum measurement to optimally discriminate between a set of quantum states, i.e., to optimize a given metric averaged over the states and the transition probabilities induced by the measurement, was first considered by Yuen, Kennedy and Lax~\cite{YKL}, who showed that the optimal measurement is one whose measurement operators satisfy a particular semi-definite program, which is described later in this paper (we use {\em optimal} measurement, {\em minimum probability of error} (MPE) measurement and {\em Yuen-Kennedy-Lax} (YKL) measurement interchangeably). For arbitrary states however, finding the solutions of the YKL semi-definite program can be computationally hard. However, \cite{Barnum} have shown that a certain measurement called the pretty good measurement (PGM) comes close to the optimal measurement for arbitrary states. Moreover, upper and lower bounds on the MPE have been recently obtained for the general problem~\cite{Tyson}. Restricting to pure states with a group symmetry makes the problem of finding the exact MPE more tractable. Helstrom considered the problem of finding the optimal measurement and the exact MPE for states that have cyclic group symmetry~\cite{Helstrom}. This was extended to arbitrary abelian groups with a transitive action by Forney and Eldar~\cite{ForneyEldar}. In fact, they showed that the pretty good measurement (PGM), also called the least squares measurement (LSM), as defined in \cite{Belavkin} and \cite{WoottersHausladen}, is optimal in this case. Later, for any non-abelian group with a transitive action on the states, Eldar et al.,~\cite{EldarMegretskiVerghese} showed that the PGM was again optimal. However, an expression for the MPE was not known. In this paper, we fill this gap using group representation theory. In \cite{Jafarizadeh}, the YKL conditions for the optimal measurement was were written in terms of equalities (rather than involving an inequality, which makes it harder to check). Recently, in \cite{Brun}, a generalization of the optimal measurement to non-projective measurements has been considered.

When the states are \emph{compound geometrically uniform} (CGU), i.e., they have multiple orbits under the group action (restricting to group actions that are permutation representations), very little is known about the structure of the optimal measurement. Solving the general CGU state discrimination problem is particularly useful for designing the optimal decoder of linear codes for classical quantum (cq) channels, i.e., for sending classical data over a quantum (such as, an optical) channel. It was recently shown that a generalization of Arikan's polar codes~\cite{Ari08} can achieve the Holevo capacity of any cq channel~\cite{Wil12a}. 


Any linear code has an automorphism group and the action of this group on the code is a permutation action and hence is a CGU set. This group action carries over, in general, to modulated code words. For example, for binary codes over a binary-phase-shift coherent-state alphabet ($|\alpha\rangle, |-\alpha\rangle$), the bit flip operation maps to the $\pi$ phase-shift operation in the modulated domain, i.e., ${\hat U}_\pi |\pm \alpha\rangle = | \mp \alpha\rangle$, where ${\hat U}_\pi = e^{i\pi {\hat a}^\dagger {\hat a}}$. Therefore, finding the optimal decoder of a set of pure states with a CGU action will not only enable finding the best performance of any coherent-state-modulated linear code, but also lend useful insight towards designing structured optical receivers to realize the optimal measurement. In \cite {Forney}, Forney gives several examples of codes whose automorphism groups have a GU action on the code space.

In this paper, we consider both the GU and the CGU actions on sets of linearly-independent pure states. In the case of GU action, we show how one can use representation theory to explicitly calculate the probability of error and the conditional probabilities. Then we consider the CGU action and show that one can reduce the problem to a set of simultaneous equations. The number of these equations depends on the sizes of the multiplicity spaces of the representation and the number of orbits. We will present examples to show the usefulness of this method when the representations have small multiplicities and few orbits. In particular, we will show examples of families of codes of length $N$ and rate $R$ (i.e., number of codewords to discriminate, $n = 2^{NR}$), such that the YKL conditions give rise to $n^2$ simultaneous equations, whereas our method requires solving a small constant number of simultaneous equations. This number comes from the dimension of the multiplicity spaces as we explain later and is independent of $N$.

This paper is organized as follows. In Section~\ref{sec:Helstrom_approach}, we describe the problem of discriminating between quantum states when the states are pure and are linearly independent. We describe the gram matrix approach to computing the YKL measurement given in Helstrom~\cite{Helstrom} along with a caveat about this approach. In Section~\ref{sec:bruteforce}, we present some examples of brute-force calculations of the MPE measurement for examples relevant to optical communication, and demonstrate why this technique is not scalable. In Section~\ref{sec:GU}, we describe the optimal measurement for GU states. This description generalizes the ones in~\cite{Helstrom, ForneyEldar} for abelian groups to non-abelian groups using group representation theory. Then in Section~\ref{sec:CGU}, we generalize the results from the previous section, and describe how one can obtain the optimal measurement for CGU states. We show how to reduce the number of simultaneous equations based on the representation of the group, from the full number of equations as specified by the YKL conditions. Then we give examples to illustrate this method, and its usefulness in the context of optical communication. Finally, in Section~\ref{sec:conclusions}, we conclude the paper with a summary and open questions.

\section{Optimal measurements for pure states}\label{sec:Helstrom_approach}
In this section, we describe two approaches to finding the optimal measurement. Suppose we are given an ensemble $\left\{p_i, |\psi_i\rangle\right\}, 1 \le i \le n$, of $n$ linearly-independent pure states and an associated prior distribution. It can be shown that when distinguishing pure states, the optimal measurement is an $n$-element projective measurement, and is unique~\cite{Helstrom}. Therefore, let us assume that the optimal measurement is given by the orthonormal basis $\left\{\ket{w_i}, 1 \le i \le n\right\}$. Now define two matrices: the matrix $M$ whose columns are the un-normalized pure-state vectors $\sqrt{p_i}\ket{\psi_i}$ and the matrix $X$ whose elements are $x_{ij}=\sqrt{p_j}\braket{w_i}{\psi_j}$. Since the states are linearly independent, each state lies in an $n$ dimensional Hilbert space and $M$ is an invertible $n\times n$ matrix. The matrix $X$ denotes the solution to the state discrimination problem since all the information about the measurement vectors can be obtained from $X$. Yuen, Kennedy and Lax showed that $X$ must satisfy the equations~\cite{YKL}:
\be\label{Eq1}
X^\dag X= \Gamma, \; {\text{and}}
\ee
\be\label{Eq2}
x_{km}x^\ast_{mm} = x_{kk}x^\ast_{mk} \,,
\ee
where $\Gamma$ is the Gram matrix of the set of states, i.e., $(\Gamma)_{ij} = \langle \psi_i | \psi_j \rangle$. In Helstrom's book~\cite{Helstrom}, it is suggested that these equations lead to the solution. However, we would like to emphasize here that these two equations alone do not give a unique solution. In \cite{YKL}, it was shown that along with the above two equations, an \emph {inequality} must be satisfied. Only when the inequality is considered, one gets a unique solution in general. However, in certain cases of interest, one can get a small set of solutions using the above two equations as we show later.

It is useful to view this in terms of the polar decomposition. The left and right polar decomposition of the matrix $M$ is given by,
\begin{equation}
M=U\sqrt{M^\dag M} = \sqrt{MM^\dag}U \,.
\label{eq:Mdef}
\end{equation}
In the above equation, $M^\dag M$ is just the Gram matrix $\Gamma$ of the set of states $\{\ket{\psi_1},\dots ,\ket{\psi_n}\}$ and $U$ is a unitary matrix. Denote $\sqrt{M^\dag M}$ by $P$. It is known that if $M$ is invertible, then $P$ and $U$ are unique, with $P$ being a positive semi-definite matrix. Clearly, $P$ satisfies $P^\dag P=\Gamma$ as does any matrix of the form $VP$, where $V$ is unitary. Since $VP$ always satisfies Eq.~\eqref{Eq1} for any unitary $V$, it is chosen so that $VP$ satisfies Eq.~\eqref{Eq2} as well. Therefore, the matrix $X$ is in general of the form $VP$ and the measurement vectors $\ket{w_i}$ are columns of the matrix $UV^\dag$. Finally, note that if the solution $X$ turns out to be such that $p_kx_{kk}=p_mx_{mm}$, then Eq.~\eqref{Eq2} becomes $x_{km}=x^\ast_{mk}$ for all $k$ and $m$, i.e., $X$ is Hermitian. Below, we will show that the above condition is satisfied when $\left\{|\psi_i\rangle\right\}$ is a GU set.

\section{Brute force calculation of the MPE measurement}\label{sec:bruteforce}

Helstrom calculated the optimal measurement for several simple examples~\cite{Helstrom}, by assuming that any symmetry in $\Gamma$ is carried over unaltered, to $X$. For example, he considered the equiprobable ternary ($n=3$) coherent state set $\left\{|-\alpha\rangle, |0\rangle, |\alpha\rangle\right\}$, $\alpha \in R$. Since the inner products, $\langle \pm \alpha | 0 \rangle = \langle 0 | \pm \alpha \rangle = e^{-|\alpha|^2/2} \equiv \kappa$, and $\langle \pm \alpha | \mp \alpha \rangle = e^{-2|\alpha|^2} = \kappa^4$, this ensemble has an `isosceles' geometry. He argues therefore,
\begin{equation}
\Gamma = \left(
\begin{array}{ccc}
1 & \kappa &\kappa\\
\kappa & 1 &\kappa^4\\
\kappa &\kappa^4 &1
\end{array}
\right) \Rightarrow
X = \left(
\begin{array}{ccc}
a & d & d\\
b & c & e\\
b & e &c
\end{array}
\right).
\end{equation}
With this assumption of symmetry for $X$, the $n^2 = 9$ simultaneous equations resulting from the YKL equality conditions Eqs.~\eqref{Eq1} and~\eqref{Eq2} reduce to $5$ simultaneous equations in $a, b, c, d, e$, which Helstrom solved (four variables were eliminated analytically, and the last one was solved for numerically using the Newton method) to obtain (all entries of) $X$ solely in terms of $\sigma$, and hence evaluated the average probability of error, $P_e = 1-(|a|^2+2|c|^2)/3$.

A rate $R$ code $\cal C$ of length $N$ codewords over a binary phase-shift keying (BPSK) alphabet $\left\{|\alpha\rangle, |-\alpha\rangle\right\}$, has $n = 2^{NR}$ codewords. Each row of the $n \times n$ Gram matrix of the codebook, $\Gamma$ is a permutation of the first row, if $\cal C$ is a linear code. Given this symmetry, the $n^2$ simultaneous equations of the YKL conditions reduce to $n$ equations, since each row of $X$ must also be the same permutation as the corresponding row of $\Gamma$. One might assert---based on Helstrom's argument on the symmetry in $\Gamma$ carrying over to $X$---that the number of distinct entries $0 < d \le n$ in each row of $X$, would be the same as the number of distinct Hamming weights ($d$) in the code. Note that $((\Gamma))_{ij} = \sigma^{2d_{ij}}$, with $\sigma = \langle \alpha | -\alpha \rangle = e^{-2|\alpha|^2}$, where we define the elements $d_{ij}$ of the code's `distance matrix' $D$ to be the Hamming weight between the $i^{\rm th}$ and the $j^{\rm th}$ codewords. The aforesaid assertion was used in Refs.~\cite{Guh11, Guh11a} to calculate the MPE measurements for some simple BPSK codes, including the first order binary Reed Muller ${\rm RM}(1,m)$ codes. 

\begin{figure}
\centering
\includegraphics[width=\columnwidth]{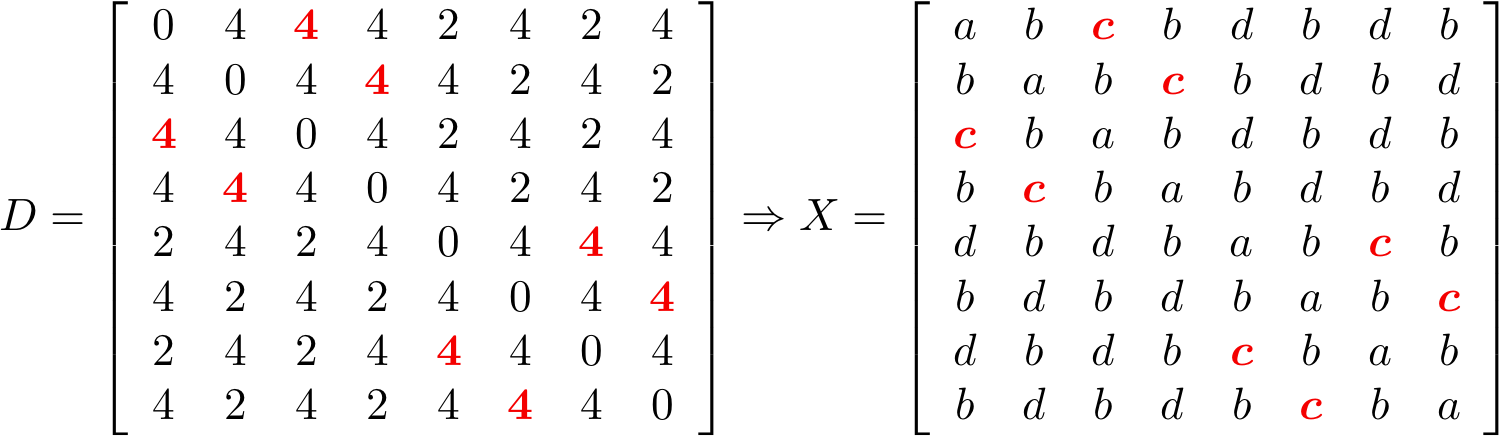}
\caption{Solving the YKL conditions numerically for the Gram matrix of a [8, 3, 2] sub code $\mathcal C$ of the BPSK ($|\alpha\rangle, |-\alpha\rangle$) Reed Muller $(r = 2, m = 3)$ [8, 7, 2] code, results in the number of distinct elements in a row of $X$ to be one more than the number of distinct Hamming distances in $\mathcal C$. Elements of the Gram matrix $\Gamma$, $\gamma_{ij} = \sigma^{d_{ij}}$, where $\sigma = \langle \alpha | -\alpha \rangle = e^{-2|\alpha|^2}$, and $d_{ij}$ are the elements of the distance matrix $D$ of the code. The $4$'s in boldface red font in each row of $D$ results in an entry at the corresponding position in the row in $X$, that is distinct from the entries in the row of $X$ corresponding to the other $4$'s in the row of $D$. For $|\alpha|^2 = 0.01$, solving this example numerically yields $a=0.54$, $b=0.294$, $c=0.263$, and $d=0.382$.}
\label{fig:numericalexample}
\end{figure}

Upon some numerical investigation, we found the aforesaid assertion to be false. We found examples of subcodes of the first order binary Reed Muller code, for which identical entries in one row of $\Gamma$ resulted in distinct entries in the corresponding row of $X$ (see Fig.~\ref{fig:numericalexample}).
This breaking of symmetry from $\Gamma \to X$ led us to look into multiple orbits, and develop the mathematics to rigorously understand the optimal measurements for CGU sets, which encapsulate all linear codes. This in turn led us to generalize and simplify the previous results for GU sets as well, which are presented in Section~\ref{sec:GU}. Our general results on MPE measurements for CGU sets are presented in Section~\ref{sec:CGU}.

\section{Geometrically uniform states and the pretty good measurement}\label{sec:GU}
We say that a set of states is {\em geometrically uniform} (GU) if there is a group $G$ acting transitively on them i.e., for every two states $\ket{\psi_i}$ and $\ket{\psi_j}$ there exists a group element $g \in G$ such that $R(g)\ket{\psi_i}=\ket{\psi_j}$, where $R$ is some representation of the group $G$. This implies that all the elements of the set are obtained from a single element, say $\ket{\psi_1}$ by the action of the group. If the states are linearly independent, then the representation of the group on the space spanned by the states is the induced representation of the trivial representation of the stabilizer subgroup of $\ket{\psi_1}$. For a state discrimination problem to be GU, one usually assumes that the priors associated with the states in the GU set are the same for all states. The PGM was proved to be optimal for cyclic groups in \cite{Helstrom}, for abelian groups in \cite{ForneyEldar} and for non-abelian groups in \cite{EldarMegretskiVerghese}. The pretty good measurement has been defined in \cite{Belavkin} and \cite{WoottersHausladen} as a measurement to discriminate between the states $\rho_i$ with priors $p_i$. The measurement operators of the PGM are given by $\Pi_i=p_i\rho^{-1/2}\rho_i\rho^{-1/2}$, where $\rho=\sum_ip_i\rho_i$. If the states are pure ($\rho_i = \ket{\psi_i} \langle \psi_i |$) and linearly independent, this measurement becomes a projective measurement. Consider the polar decomposition of the matrix $M$ (defined in Eq.~\eqref{eq:Mdef}).
Observe that $\rho=MM^\dag$ and so $U=\rho^{-1/2}M$. The columns of $U$ form the measurement basis of the PGM. From the left polar decomposition, notice that the columns of $U$ are also the measurement basis if the solution matrix $X$ coincides with $P=\sqrt{M^\dag M}$. Since $P$ is Hermitian, it would be the solution of Eqs.~\ref{Eq1} and \ref{Eq2} if, in addition, all the diagonal elements of $P$ are equal (since the priors are equal). To see that this is true for geometrically uniform states, observe that $\rho$ commutes with the representation $R$ and $x_{kk}=\braket{w_k}{\psi_k}=\bra{w_1}R(g)^{-1}R(g)\ket{\psi_1}=\braket{w_1}{\psi_1}=x_{11}$.

Now we describe the measurement using non-abelian group representation theory along the lines of \cite{Helstrom, ForneyEldar} where it was done for abelian groups. In accordance with the action of the group, we have $\ket{\psi_{i}}=U(g_{i})\ket{\psi_{1}}$ for any $i$. We assume that the priors $p_{i}$ are all the same. Any transitive permutation action on a linearly independent set is an induced representation. We pick a base point, say $\ket{\psi_{1}}$ and with respect to this point, there is a subgroup $G_{0}$ of $G$ which stabilizes $\ket{\psi_{1}}$. The representation on the vector space spanned by $\ket{\psi_{i}}$ is the induced representation of the trivial representation of $G_0$ to $G$. If the set of states is $S$, then we have that $|S|=|G|/|G_0|$.

The Yuen, Kennedy, Lax conditions are
\begin{align}
&\Upsilon - p_{i}\psi_{i} \geq 0, \nonumber \\
&(\Upsilon - p_{i}\psi_{i})\Pi_{i} = 0, \;{\text{and}} \nonumber \\
&\Upsilon = \sum_{i} p_{i} \psi_{i}\Pi_{i} = \sum_{i} p_{i} \Pi_{i}\psi_{i} \,,
\end{align}
where $\psi_{i}=\ket{\psi_{i}}\bra{\psi_{i}}$. Since the optimal measurement basis for a GU set is also GU, it is easy to see that we only need the equations where in the first two $i=1$.

Let the optimal measurement basis be given by $\{\ket{w_{i}}\}$ which are also GU under the $G$ action (and let $\Pi_{i} = \ket{w_{i}}\bra{w_{i}}$). Therefore we have
\[
\Upsilon=\frac{1}{|G|}\sum_{g\in G} U(g) \psi_{1}\Pi_{1} U(g^{-1}).
\]
Now let this representation consist of irreducible representations $\lambda$ with multiplicity $m_\lambda$. Consider the Fourier basis $\ket{\lambda,m,k}$ where $\lambda$ labels the irreducible representation, $m$ its multiplicity and $k$ its representation space whose dimension is denoted $d_\lambda$. The matrices $U(g)$ are block diagonal in this basis and therefore the operator $\Upsilon$ is also block diagonal by Schur's lemma. In order to find the probability of error, we need to access to an arbitrary matrix element of $\Upsilon$ inside the blocks. Following Helstrom \cite{Helstrom} (who worked this out for cyclic groups), we have
\begin{align}
&\bra{\lambda,m,k}\Upsilon\ket{\lambda^\prime,m',k'} =\nonumber \\
& \frac{1}{|G|}\sum_{g} \bra{\lambda,m,k} U(g) \psi_1 \Pi_1  U(g^{-1}) \ket{\lambda',m',k'}\,.
\end{align}
We denote $\psi_1$ and $\Pi_1$ as $\psi$ and $\Pi$ respectively. The action of any $U(g)$ on the state $\ket{\lambda,m,k}$ is given as follows
\[
U(g)\ket{\lambda,m,k}=\sum_{k'}\lambda(g)_{k',k}\ket{\lambda,m,k'}\,,
\]
where $\lambda(g)_{k',k}$ is the $k',k$ matrix entry of the irreducible representation $\lambda$.
Using this we get
\begin{align}
&\bra{\lambda,m,k}\Upsilon\ket{\lambda^\prime,m',k'} = \nonumber \\
&\frac{1}{|G|}\sum_{g,l,l'} \lambda^\ast(g^{-1})_{k,l}\lambda'(g^{-1})_{k',l'} \bra{\lambda,m,l} \psi \Pi \ket{\lambda',m',l'}\,.
\end{align}
Using the orthogonality relations among matrix entries of irreducible representations, we obtain
\[
\bra{\lambda,m,k}\Upsilon\ket{\lambda^\prime,m',k'} =
\frac{\delta_{\lambda,\lambda^\prime}\delta_{k,k^\prime}}{d_\lambda}\sum_l \bra{\lambda,m,l} \psi\Pi \ket{\lambda,m^\prime,l}.
\]
One can check that $\Upsilon$ is block diagonal with the blocks given by multiplicity spaces.
Let $\phi=(1/|S|)\psi$. Now using the YKL equations, inside these invariant spaces, we see that
\begin{align}\label{Eq:Upsilon_c}
&\bra{\lambda,m,k} (\Upsilon\Pi - \phi\Pi) \ket{\lambda',m',k'} = 0\nonumber \\
&=\Upsilon_{\lambda,m}\gamma_{\lambda,m,k} \gamma^{\ast}_{\lambda',m',k'} - x_{\lambda,m,k}\gamma^{\ast}_{\lambda',m',k'}\braket{\phi}{w}=0\,,
\end{align}
where $x_{\lambda,m,k}=\braket{\lambda,m,k}{\phi}$ and $\gamma_{\lambda,m,k}=\braket{\lambda,m,k}{w}$.
In order to find the optimal measurement, we need to solve for $\gamma$. Suppose that $\gamma_{\lambda,m,k}=x_{\lambda,m,k}/c_{\lambda,m}$, where we need to solve for $c_{\lambda,m}$. We have from Eq. \ref{Eq:Upsilon_c} that $\Upsilon_{\lambda,m} = c_{\lambda,m} \braket{\phi}{w}$, if $\gamma_{\lambda,m,k}\neq 0$.
We have that
\begin{align}
&\Upsilon_{\lambda,m} = \frac{|S|}{d_\lambda}\sum_{k} x_{\lambda,m,k} \gamma^{\ast}_{\lambda,m,k} \braket{\phi}{w} \nonumber \\
&=\frac{|S|}{d_\lambda} \frac{\sum_{k}|x_{\lambda,m,k}|^2}{c_{\lambda,m}} \braket{\phi}{w} \,.
\end{align}
Using the above two equations for $\Upsilon_{\lambda,m}$ we get that
\[
\frac{|S|\sum_k|x_{\lambda,m,k}|^2}{d_\lambda(c_{\lambda,m})^2} = 1\,.
\]
The solution of the above equation is
\[
c_{\lambda,m} = \sqrt{\frac{\sum_k|S||x_{\lambda,m,k}|^2}{d_\lambda}}\,.
\]
We also need to check that $\Upsilon - \phi \geq 0$ for this solution. In order to do this, let $\ket{\mu}$ be an arbitrary normalized state. Then the above equation becomes $\bra{\mu}(\Upsilon - \phi)\ket{\mu} \geq 0$. The left hand side can be written in the Fourier basis as
\begin{align}
&\sum_{\lambda,m,k}|\mu_{\lambda,m,k}|^2\Upsilon_{\lambda,m} - \left|\sum_{\lambda,m,k} x_{\lambda,m,k} \mu^\ast_{\lambda,m,k}\right|^2 \nonumber \\
&=\sum_{\lambda,m,k}|\mu_{\lambda,m,k}|^2 c_{\lambda,m} \sum_{\lambda',m',k'}\frac{|x_{\lambda',m',k'}|^2}{c_{\lambda',m'}} \quad -\nonumber \\
&\left|\sum_{\lambda,m,k} x_{\lambda,m,k} \mu^\ast_{\lambda,m,k}\right|^2  \,.
\end{align}
Now consider the second half of the above expression
\begin{align}
&\left|\sum_{\lambda,m,k} x_{\lambda,m,k} \mu^\ast_{\lambda,m,k}\right|^2 = \left|\sum_{\lambda,m,k} \sqrt{c_{\lambda,m}}\gamma_{\lambda,m,k} \sqrt{c_{\lambda,m}}\mu^\ast_{\lambda,m,k}\right|^2 \nonumber \\
&\leq \sum_{\lambda,m,k}|\mu_{\lambda,m,k}|^2 c_{\lambda,m} \sum_{\lambda',m',k'}\frac{|x_{\lambda',m',k'}|^2}{c_{\lambda',m'}} \,,
\end{align}
where the last line was obtained through Cauchy-Schwartz. This shows that the $\gamma$ are the solutions. We can assume that the basis of the multiplicity space is picked in such a way that $x_{\lambda,m,k}$ is non-zero for only one $m$. Then $c_{\lambda,m}=\sqrt{\Tr(P_\lambda\psi)/d_\lambda}$ and
\[
\ket{w}=\sum_{\lambda,m,k} \ket{\lambda,m,k}\frac{x_{\lambda,m,k}}{c_{\lambda,m}}=\sum_\lambda \sqrt{\frac{d_\lambda}{|S|}}\frac{P_\lambda\ket{\psi}}{\sqrt{\bra{\psi}P_\lambda\ket{\psi}}}\,,
\]
where $P_\lambda$ is the projector onto the isotypic space $\lambda$.

We now calculate the probability of success using this expression. The probability of success is given by $P_s=|\braket{w}{\psi}|^2$. This can be written as
\[
P_s=\left|\sum_\lambda \sqrt{\frac{d_\lambda}{|S|}}\sqrt{\bra{\psi}P_\lambda\ket{\psi}}\right|^2\,.
\]
For any group with a representation $U$, an expression for $P_\lambda$ is given by (for a character $\chi_\lambda$)
\[
P_\lambda = \frac{d_\lambda}{|G|}\sum_g \chi_\lambda(g^{-1}) U(g).
\]
One can simplify the expression $\bra{\psi}P_\lambda\ket{\psi}$ as follows.
\be
\bra{\psi}P_\lambda\ket{\psi}=\frac{d_\lambda}{|G|}\sum_{g\in G} \chi_\lambda(g^{-1})\bra{\psi}U(g)\ket{\psi}\,.
\ee
But we have $\bra{\psi}U(g)\ket{\psi}=\bra{\psi}U(g_1gg_2)\ket{\psi}$ for all $g_1,g_2\in G_0$, where $G_0$ is the stabilizer group of $\ket{\psi}$ i.e., the subgroup of $G$ such that $U(g)\ket{\psi}=\ket{\psi}$, $\forall g\in G_0$. This means
\be
\bra{\psi}P_\lambda\ket{\psi}=\frac{d_\lambda}{|G|}\sum_{i,g\in C_i} \chi_\lambda(g^{-1})\bra{\psi}U(g)\ket{\psi}\,,
\ee
where $i$ is a sum over $(G_0,G_0)$ double coset representatives and $C_i$ is the double coset. This sum can be further simplified to
\be
\bra{\psi}P_\lambda\ket{\psi}=\frac{d_\lambda}{|G|}\sum_{i}\bra{\psi}U(g_i)\ket{\psi}\sum_{g\in C_i} \chi_\lambda(g^{-1})\,,
\ee
where $g_i$ is the double coset representative of the double coset $C_i$. This can further be written as 
\be
\bra{\psi}P_\lambda\ket{\psi}=\frac{d_\lambda}{|S|}\sum_{i}\chi_\lambda(C_i)\bra{\psi}U(g_i)\ket{\psi}\,,
\ee
where $\chi_\lambda(C_i)=(1/|G_0|)\sum_{g\in C_i}\chi_\lambda(g^{-1})$. This gives an explicit formula for the probability of success. This means that if we can find the sum of the character values of the elements of a double coset easily, then we can obtain the formula for the probability of success. In the next subsection, we show how to do this for a specific case.

\subsection{GU example from optical communication}

Let us first consider the example of the $N$-ary optical pulse position modulation (PPM), which has $N$ codewords each consisting of $N$ modes, only one of which is excited in a coherent-state pulse $|\alpha\rangle$, $\alpha \in {\mathbb R}$, where ${\bar n} = |\alpha|^2$ is the mean photon number of the pulse. Each row of the Gram matrix has two distinct entries: (a) one diagonal entry, $\langle \psi_i | \psi_j \rangle = 1$, and (b) $N-1$ entries corresponding to distinct codewords, $\langle \psi_i | \psi_j \rangle = \kappa^2$, where $\kappa \equiv \langle 0 | \alpha \rangle = e^{-{\bar n}/2}$. The PPM state set is clearly GU under cyclic group action, for which Helstrom's calculation of the error probability~\cite{Helstrom} can be applied to obtain $P_e^{\rm MPE} = \frac{N-1}{N^2}\left[\sqrt{1+(N-1)\kappa^2} - \sqrt{1-\kappa^2}\right]^2$. This was also independently obtained earlier by Liu in~\cite{Liu68}.

The standard receivers employed in optical communication are homodyne detection, heterodyne detection, and direct detection. The optimal standard receiver measurement for demodulating PPM is direct detection. The quantum-noise-limited direct detection measurement is realized by an ideal photon-number resolving (PNR) detector, which is a projective measurement on the photon number basis $\left\{|0\rangle, |1\rangle, \ldots\right\}$. In this basis, a coherent state $|\alpha\rangle = e^{-|\alpha|^2/2}\sum_{k=0}^\infty \frac{\alpha^k}{k!}|k\rangle$. An ideal PNR detector will successfully discriminate the PPM codewords if either (a) the pulse position in the codeword successfully generates a `click' (which happens with probability $1 - e^{-{\bar n}}$), or (b) if the pulse fails to generate a click but the receiver still chooses the correct codeword purely by chance. Assuming all the codewords are equally likely, the average probability of error, $P_e^{\rm PNR} = \frac{N-1}{N}e^{-{\bar n}}$. Dolinar showed that in the `high photon number' regime ($Ne^{-{\bar n}} \ll 1$ to be precise), $P_e^{\rm PNR} \sim e^{-{\bar n}}$, whereas $P_e^{\rm MPE} \sim e^{-2{\bar n}}$~\cite{Dol76}. Thus, the MPE has a factor of two higher error exponent compared with quantum-noise-limited direct detection, in the high-photon number limit. Even though the design of a structured optical receiver that can exactly attain $P_e^{\rm MPE}$ at any value of ${\bar n}$ remains unknown, there are receiver structured known---the {\em conditional pulse nulling} (CPN) receiver for discriminating PPM codewords~\cite{Dol76, Guh11a}, and the {\em sequential waveform nulling} (SWN) receiver which works for discriminating {\em any} $N$ coherent state codewords~\cite{Nai14}---both of which can attain the optimal error exponent (i.e., that of the MPE measurement) in the high photon number limit.

Let us recall that the PGM was proved to be the optimal (MPE) measurement for: (a) cyclic groups in \cite{Helstrom}, for (b) abelian groups in \cite{ForneyEldar}, and for (c) non-abelian groups in \cite{EldarMegretskiVerghese}. For the cases dealt in (b) and (c), no systematic method to calculate the minimum error probability was given. Hence, we consider next an example from optical communication that truly demonstrates the power of our method since the group involved is non-abelian, and hence no systematic method to calculate the MPE (other than by brute-force evaluation) is known.  Two-pulse PPM is a modulation constellation containing $\binom{N}{2}$ codewords. Each codeword consists of $N$ modes, whose state is a tensor product of coherent state $|\alpha\rangle$ in two of the $N$ modes, and vacuum ($|0\rangle$) in the remaining $N-2$ modes. Each row of the Gram matrix therefore has three distinct entries: (a) one diagonal entry, $\langle \psi_i | \psi_j \rangle = 1$, (b) $\binom{N-2}{2}$ entries corresponding to codeword pairs both of whose pulses are in non-overlapping modes, i.e., $\langle \psi_i | \psi_j \rangle = \kappa^4$, where $\kappa \equiv \langle 0 | \alpha \rangle = e^{-|\alpha|^2/2}$, and finally, (c) remaining entries corresponding to codeword pairs one of whose pulses are in the same mode, i.e., $\langle \psi_i | \psi_j \rangle = \kappa^2$.

We assume some familiarity with the representation theory of the symmetric group. The symmetric group $S_N$ acts on this set of states and in this representation of $S_N$, there are three irreducible representations. The trivial, standard and the a third irrep whose Young diagram has two rows with two boxes in the second row (label them $a, b, c$ respectively). In this case, $G_0$ is $S_2\times S_{N-2}$ and the double coset representatives are $e,(1,3)$ and $(1,3)(2,4)$ (label these double cosets $C_0$, $C_1$ and $C_2$ respectively). The double coset sums turn out to be
\begin{align}\label{eq:double_coset_sums}
&\chi_a(C_0)=1\,,\chi_a(C_1)=2(N-2)\,,\chi_a(C_2)={N-2\choose 2}\,,\nonumber \\
&\chi_b(C_0)=1\,,\chi_b(C_1)=N-4\,,\chi_b(C_2)=-(N-3)\,,\nonumber \\
&\chi_c(C_0)=1,\chi_c(C_1)=-2\,,\chi_c(C_2)=1\,.
\end{align}
where $\chi_\lambda(C_i)=(1/|G_0|)\sum_{g\in C_i}\chi_\lambda(g^{-1})$. With this the probability of success can then be easily calculated to be
\begin{align}
&P_s^{\rm MPE}=\bigg|\frac{2}{N(N-1)}\bigg[\sqrt{1+\chi_a(C_1)\kappa^2+\chi_a(C_2)\kappa^4}\nonumber\\
&+(N-1)\sqrt{1+\chi_b(C_1)\kappa^2+\chi_b(C_2)\kappa^4}\nonumber\\
&+\left({N\choose 2}-N\right)\sqrt{1+\chi_c(C_1)\kappa^2+\chi_c(C_2)\kappa^4}\bigg]\bigg|^2\,.
\end{align}
When $\kappa=0$, it can be seen that $P_s=1$. This corresponds to the states being orthogonal to each other and the PGM is just the measurement in that basis (and it always succeeds).

An ideal PNR detector will successfully discriminate the codewords if either (a) both pulse positions generate a `click' (each of which happens with probability $1 - |\langle 0 | \alpha \rangle|^2 = 1-\kappa^2$), (b) only one of the pulses generate a click and a random guess among the $N-1$ remaining pulse positions yields the second pulse position correctly by chance, or (c) neither of the pulses generate clicks but a random guess among all $\binom{N}{2}$ codewords yields the correct choose by chance. The success probability is therefore given by,
\begin{align}
&P_s^{\rm PNR}=\left(1-\kappa^2\right)^2 + \frac{2\left(1-\kappa^2\right)\kappa^2}{N-1}
+ \frac{\kappa^4}{\binom{N}{2}}.
\end{align}
\begin{figure}
\centering
\includegraphics[width=\columnwidth]{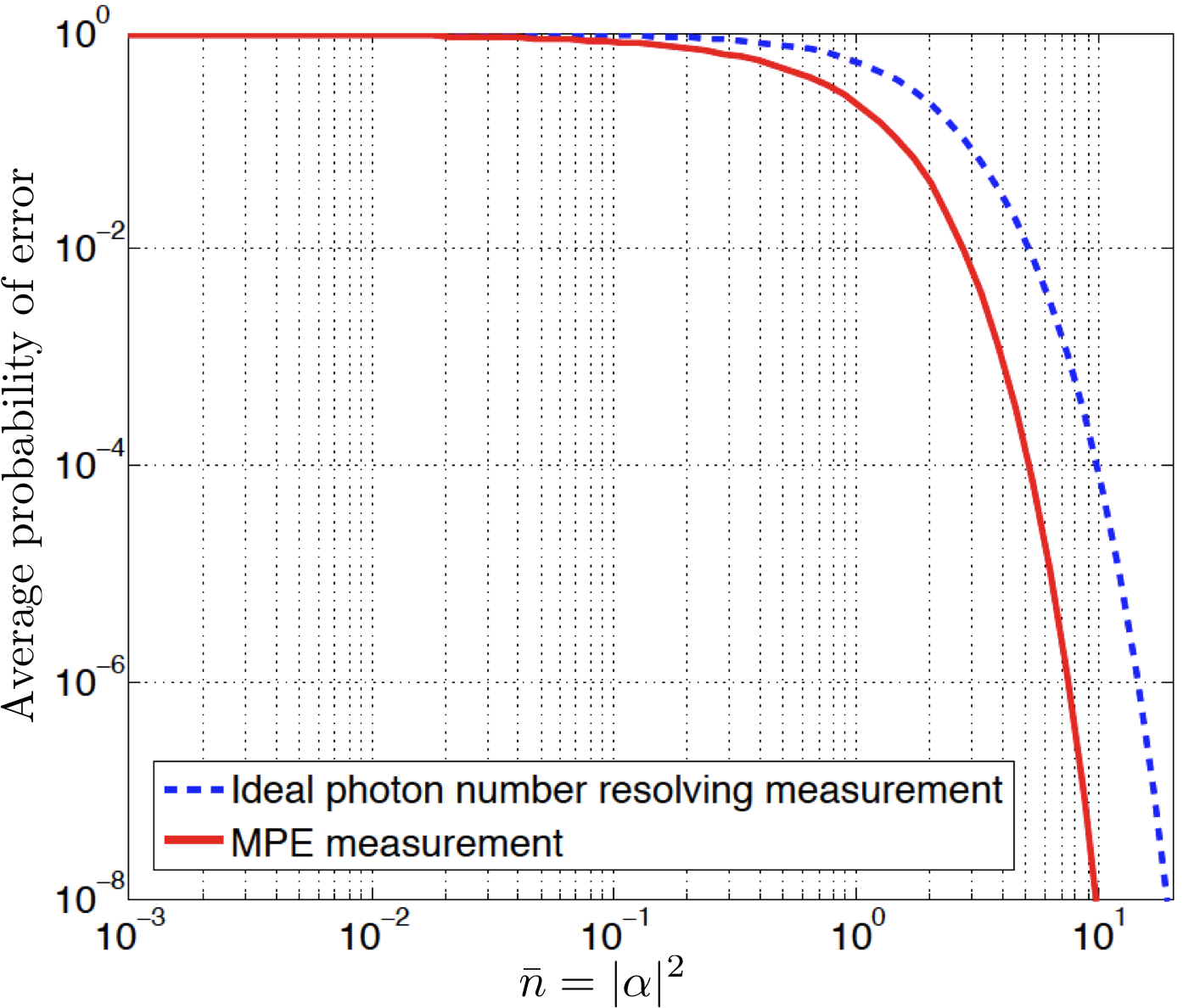}
\caption{(Color online) The probability of error of discriminating $\binom{N}{2}$ two-pulse-PPM coherent-state codewords, for $N=8$, plotted as a function of the mean photon number ${\bar n} = |\alpha|^2$. The blue (dashed) plot corresponds to the error probability achievable by a quantum-noise-limited PNR measurement, whereas the red (solid) plot is the minimum probability of error (MPE) achievable by the optimal measurement allowable by quantum mechanics.}
\label{fig:GU_example}
\end{figure}

The probabilities of error $P_e^{\rm MPE} = 1 - P_s^{\rm MPE}$, and $P_e^{\rm PNR} = 1 - P_s^{\rm PNR}$ are plotted as a function of the mean photon number ${\bar n} = |\alpha|^2$ in Fig.~\ref{fig:GU_example}. Interestingly, numerical evaluation of these error probabilities for for two-pulse PPM show that, just like PPM, in the high photon number limit ($Ne^{-{\bar n}} \ll 1$), $P_e^{\rm PNR} \sim e^{-{\bar n}}$, whereas $P_e^{\rm MPE} \sim e^{-2{\bar n}}$.

It is worth noting that the SWN receiver~\cite{Nai14}---which can be built in principle using simple linear-optic components and single photon detectors---can attain the optimal (MPE measurement's) error exponent in the high ${\bar n}$ limit for discriminating any $N$ coherent state waveforms. Hence for two-pulse PPM, $P_e^{\rm SWN} \sim e^{-2{\bar n}}$. An optical receiver structure to exactly attain $P_e^{\rm MPE}$ at any finite $\bar n$ is significantly more complicated, and requires truly non-classical (entangling) operations within the receiver~\cite{daS13}. 

\section{Optimal measurement for CGU sets}\label{sec:CGU}

In this section, we describe how to obtain the optimal measurement for CGU state sets. We use the Helstrom description of the problem of finding the optimal measurement for pure states i.e., by viewing it as the solution of a set of simultaneous equations. However, as we pointed out earlier, one need not obtain a unique solution. With every obtained solution, we have to check the third condition to find the right one. We again resort to representation theory to simplify the equations and obtain far fewer equations (in many practical cases of interest). We begin by recalling the set of simultaneous equations which give the solution given above in Eqs.~\ref{Eq1} and \ref{Eq2}. $X^\dag X= \Gamma, \quad x_{km}x^\ast_{mm} = x_{kk}x^\ast_{mk}$ where $\Gamma$ is the Gram matrix of the set of states. Since we have CGU symmetry in the problem, the Gram matrix is symmetric about a group $G$ and its representation $U(g)$. Suppose this representation decomposes into irreducible spaces $\lambda$ of dimension $d_\lambda$ with multiplicity $m_\lambda$ (as before). Then the Gram matrix is block diagonal in this basis with a block corresponding to each irreducible representation. Denote these blocks as $\Gamma'_\lambda$. These blocks also have a special structure where they are identity in the representation space i.e., $\Gamma'_\lambda=\Gamma_\lambda\otimes I_{d_\lambda}$. We note that the solution $X$ would have the same block diagonal decomposition since it commutes with the same representation of $G$. Then we have that inside an isotopic space $X_\lambda^\dag X_\lambda=\Gamma_\lambda$. Notice that these matrices are of dimension $m_\lambda\times m_\lambda$. Therefore, if the multiplicity spaces are small, then this task is easy. Now, solving for such an $X_\lambda$ can be done only up to a unitary since for any solution $X_\lambda$, $U_\lambda X_\lambda$ is also a solution, where $U_\lambda$ is an arbitrary unitary operator. In order to find a solution, we need to use the set of equations in Eq.~\ref{Eq2}.

\subsection{CGU example from optical communication}

\begin{figure}
\centering
\includegraphics[width=\columnwidth]{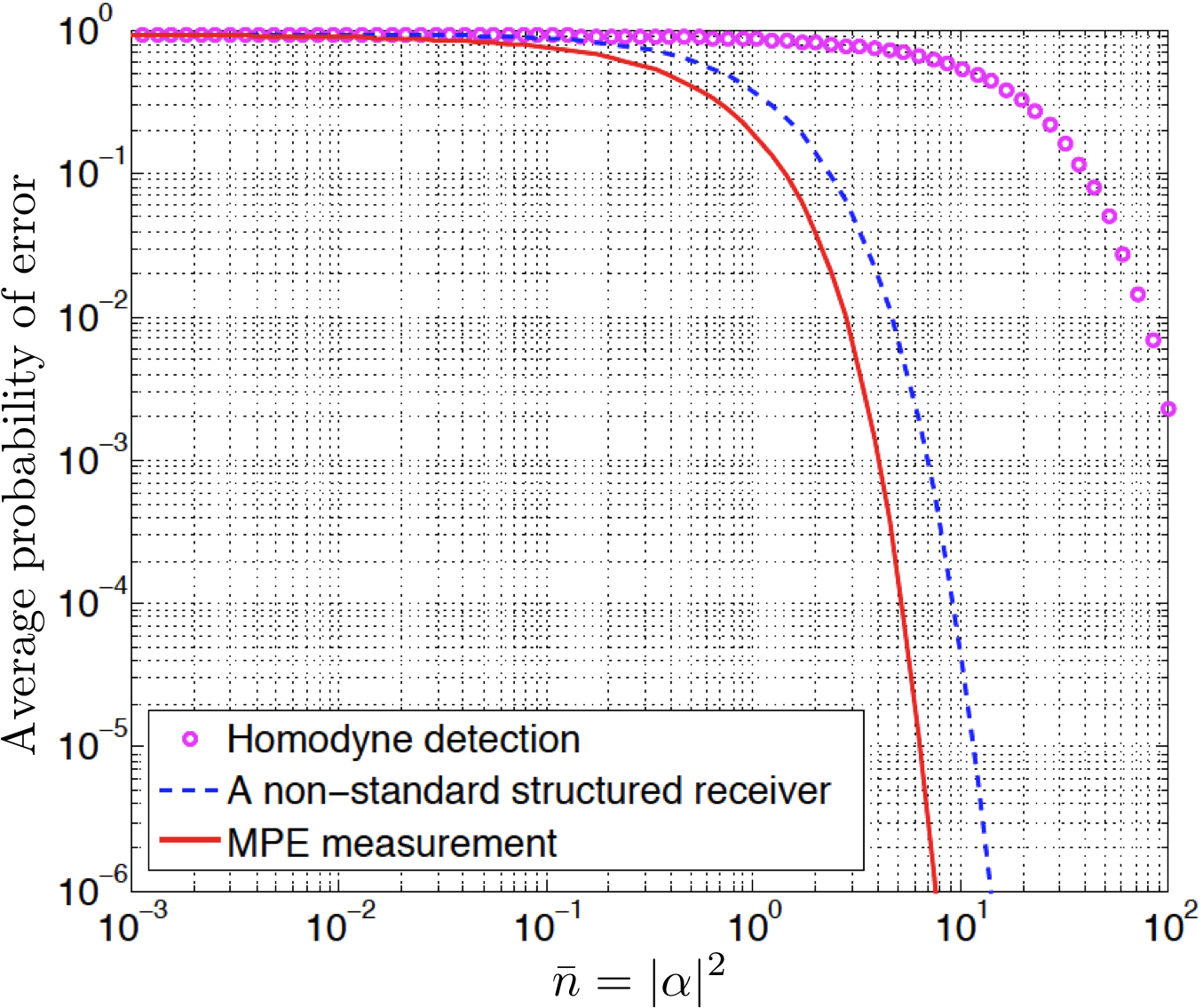}
\caption{(Color online) The probability of error of discriminating $2N$ binary-phase-coded PPM codewords, for $N=8$, plotted as a function of the mean photon number ${\bar n} = |\alpha|^2$. The magenta line (circles) correspond to the error probability achievable by an ideal homodyne detection measurement, the blue (dashed) plot is the error probability achievable by a non-standard yet structured receiver described in the text, and the red (solid) plot is the minimum probability of error (MPE) achievable by the optimal measurement allowable by quantum mechanics.}
\label{fig:CGU_example}
\end{figure}
CGU state sets are of particular importance in optical communication. This is because all linear codes have CGU symmetry, and explicit linear codes (viz., quantum polar codes) are known to achieve the quantum (Holevo) limit of the classical communication capacity over any quantum channel~\cite{Wil12a}, including the lossy-noisy bosonic channel~\cite{Guh12}. We now illustrate our method explained above with a non-trivial CGU example relevant to optical communication. Consider a modulation code comprising $n = 2N$ codewords, where each codeword is an $N$-mode pure states. Further, $N$ of the $2N$ codewords comprise a PPM set with a coherent state $|\alpha\rangle$ in the respective pulse positions, while the remaining $N$ codewords comprise a PPM set with a coherent state $|\beta\rangle$ in the respective pulse positions. This set clearly is CGU under the cyclic group action. 
Therefore, in the Fourier basis, the Gram matrix has $N$ $2\times 2$ blocks. The first block is
\[
\mx{
1+(N-1)e^{-|\alpha|^2} && C\\
C && 1+(N-1)e^{-|\beta|^2} 
}\,,
\]
where $C=\exp(\alpha\beta^\ast-\frac{|\alpha|^2+|\beta|^2}{2})$ $+ (N-1)\exp(\frac{|\alpha|^2+|\beta|^2}{2})$. The other blocks are
\[
\mx{
1-N(N-1)e^{-|\alpha|^2} && D\\
D && 1-N(N-1)e^{-|\beta|^2}
}\,,
\]
where $D=\exp(\alpha\beta^\ast-\frac{|\alpha|^2+|\beta|^2}{2})$ $- N(N-1)\exp(\frac{|\alpha|^2+|\beta|^2}{2})$. Suppose that $X$ has entries $x_{11},\dots x_{22}$ for the first block and $y$'s for the other $2\times 2$ blocks, then we obtain the equations
\begin{align}
&(x_{12} + (N-1) y_{12}) (x_{22}^\ast + (N-1) y_{22}^\ast) =\nonumber  \\
&(x_{11} + (N-1) y_{11})(x_{21}^\ast + (N-1)y_{21}^\ast) \text{  and}\nonumber\\
&(x_{12} - y_{12}) (x_{22}^\ast - y_{22}^\ast) = \nonumber\\
&(x_{11} - y_{11})(x_{21}^\ast - y_{21}^\ast)
\end{align}
For many such orbits, these equations can be generalized easily. After solving these equations, one obtains the following optimal average probability of success,
\begin{align}
&P_s^{\rm MPE} =\bigg(\sqrt{1+(N-1)e_1 + e_2 +(N-1)e_3} \nonumber\\
&+\sqrt{1+(N-1)e_1 - e_2 -(N-1)e_3} \nonumber\\
&+(N-1)\sqrt{1-e_1+e_2-e_3} \nonumber\\
&+(N-1)\sqrt{1-e_1-e_2+e_3}\bigg)^2\frac{1}{4N^2},\label{eq:CGUformula}
\end{align}
where $e_1=\exp(-|\alpha|^2)$, $e_2=\exp(-\frac{1}{2}|\alpha-\beta|^2)$ and $e_3=\exp(-\frac{1}{2}(|\alpha|^2+|\beta|^2))$.

In order to compare the minimum probability of error $P_e^{\rm MPE}=1-P_s^{\rm MPE}$ to the average error probability achievable by standard optical receivers, let us consider the case of $\beta = -\alpha \in {\mathbb R}$ in above. The modulation format thus obtained is known as (binary) phase-coded PPM, or PCPPM. The best standard optical receiver to decode the $2N$ codewords is homodyne detection. Direct detection can only discriminate between the $N$ pulse positions, but cannot discern the phase. This is because the mean photon number in the pulse, $|\beta |^2 = |\alpha|^2 = {\bar n}$ is the same for each phase. Ideal homodyne detection of a coherent state $|\alpha\rangle$ generates a Gaussian distributed random variable with mean $\alpha$ and variance $\frac{1}{4}$. Therefore, homodyne detection of all the pulse positions generates $N$ statistically-independent real-valued random variables, of which one random variable $X \sim {\cal N}(\pm \alpha, \frac{1}{4})$, and $N-1$ i.i.d. random variables $Z_i \sim {\cal N}(0, \frac{1}{4})$. The receiver first chooses the pulse position as the one the homodyne output corresponding to which has the largest absolute value. Then it chooses the phase based on the sign of the real-valued homodyne output for that pulse position. The success probability is thus given by,
\begin{equation}
P_s^{\rm hom} = {\rm Pr}\left[|X| > {\rm max}_{1 \le i \le N-1}\left\{|Z_i|\right\}\right] \left[1 - \frac{1}{2}{\rm erfc}\left(\sqrt{2{\bar n}}\right)\right], \nonumber
\end{equation}
where $|X|$ and $|Z_i|$ have folded normal distributions. We evaluated this numerically for $N=8$ and plotted the error probability $1 - P_s^{\rm hom}$ as a function of $\bar n$, in Fig.~\ref{fig:CGU_example} (see magenta circles). We also calculated and plotted the MPE using our results (Eq.~\eqref{eq:CGUformula}) as a function of $\bar n$ (see solid red line). Our MPE calculation helps show how inferior the performance is (compared with the optimal measurement) for a PCPPM modulation for the best standard optical receiver choice.

Now we consider a non-standard, yet intuitive and structured optical receiver, to detect the PCPPM codewords: The coherent-state codeword impinges a photon counting receiver. If no click is registered over all the $N$ modes (which happens with probability $p_0 = e^{-{\bar n}}$), the receiver chooses randomly between the $2N$ codewords. The first photon click must identify the pulse slot correctly. The photon arrivals within the time slot containing the pulse (in state $|\alpha\rangle$ or $|-\alpha\rangle$) are Poisson distributed. As soon as the first click arrives---the time of arrival of which is random (exponentially distributed)---the remainder of the pulse (which is in a coherent state $|\beta\rangle$ or $|-\beta\rangle$ with $|\beta|^2 < |\alpha|^2$) is switched into a Dolinar receiver~\cite{Dol76}, which identifies the phase correctly with an error probability $\frac{1}{2}\left[1-\sqrt{1-e^{-4|\beta|^2}}\right]$. It is straightforward to show that the eventual probability of error attained by this structured receiver is given by:
\begin{equation}
P_e^{\rm structured} = p_0\left[\frac{2N-1}{2N}\right] + \frac{1-p_0}{2} - \int_{p_0}^1 \sqrt{1-\left(\frac{p_0}{x}\right)^4}{\rm d}x \nonumber,
\end{equation}
which is plotted in Fig.~\ref{fig:CGU_example} (see dashed blue line). A numerical evaluation of the error probabilities for the MPE and the structured receivers show that, in the high photon number limit ($Ne^{-{\bar n}} \ll 1$), $P_e^{\rm structured} \sim e^{-{\bar n}}$, whereas $P_e^{\rm MPE} \sim e^{-2{\bar n}}$. The SWN receiver~\cite{Nai14} can attain the MPE measurement's error exponent in the high ${\bar n}$ limit for discriminating any $N$ coherent state waveforms, and hence applies to PCPPM as well. A simple structured receiver to exactly attain $P_e^{\rm MPE}$ at any finite $\bar n$ is not known, but our calculation of the optimal measurement for CGU sets allows one to use the general receiver concept in Ref.~\cite{daS13}, which---despite requiring complicated non-classical operations within the receiver---can in principle attain the MPE exactly at all $\bar n$.

\section{Conclusions}\label{sec:conclusions}

We have developed a new and compact interpretation---using group representation theory---of the minimum probability of error (MPE) measurement for distinguishing a set of {\em geometrically uniform} pure quantum states---states that form a single orbit under the group action, i.e., a transitive action. We also give a representation theoretic proof that the pretty good measurement, or equivalently the least squares measurement is the optimal (MPE) measurement for a GU set of states. More importantly, this representation theoretic framework gives explicit formulae for the minimum probability of error. This is useful in comparing the relative performance of various receivers. Using the same framework, we then extended our analysis to construct optimal measurements for {\em compound geometrically uniform} (CGU) state sets, which are states that form multiple orbits under the group action. CGU sets appear in many practical problems, particularly in transmitting classical data over a (quantum) optical channel. All linear codes formed using pure-state modulation constellations, which are known to achieve the quantum (Holevo) limit to the capacity of optical communication, are CGU sets. We showed how to compute the optimal measurement for CGU sets by reducing the problem to solving a few simultaneous equations. The number of equations depends on the sizes of the multiplicity spaces of irreducible representations. For many group representations (such as those of several practical good linear codes), this is a {\em lot} more tractable than solving large semi-definite programs, in order to solve---by brute force---the Yuen-Kennedy-Lax conditions~\cite{YKL} for determining optimal measurements for discriminating an arbitrary set of pure states with given pairwise inner products. We showed one example each of the evaluation of optimal measurements for GU and CGU states, respectively.

It is known that coherent-state (laser light) modulation is sufficient to achieve the {\em Holevo capacity}, the ultimate rate of reliable classical communication over a lossy-noisy optical channel~\cite{Gio04, Gio14}. It is also known that linear codes (over an underlying coherent-state modulation) along with optimal measurements---which are CGU sets by definition---suffice to attain the Holevo capacity~\cite{Guh12}. There is however a significant gap between the Holevo capacity and the Shannon capacity of the optical channel attainable by conventional optical receivers, viz., homodyne, heterodyne, and direct-detection receivers~\cite{Tak14}, and the gap widens in the low photon number regime~\cite{Chu11}. It would be interesting to investigate explicit finite blocklength code families with good symmetry properties, whose rate performance along with the respective optimal measurements---calculated exactly by the general method we developed---bridges the aforesaid capacity gap. It would also be interesting to develop rigorous foundations for translating the optimal CGU measurement, to an algorithmic design of structured optical receivers built using a small universal set of known optical components and ancilla states, that can implement the optimal measurement on any given linear code. In the high photon number regime on the other hand, heterodyne detection is known to be asymptotically capacity optimal. However, in the high photon number regime, the improvement attained by the MPE measurement (over conventional optical receivers) in the error exponent in discriminating symbols of a modulation constellation (as seen in our GU and CGU examples in this paper, and also in Ref.~\cite{Nai14}) translates to a superior finite blocklength rate achievable by the MPE measurement, even though heterodyne detection is capacity-optimal in this regime~\cite{Tan15}. This suggests that translating our development in this paper to an algorithmic design of structured MPE-attaining optical receivers, may also have a benefit in the high photon number transmission regime, in terms of the finite-codelength rate performance.

\begin{acknowledgments}
This material is based upon work supported by the Defense Advanced Research Projects Agency's (DARPA) Information in a Photon (InPho) program, under Contract No. HR0011-10-C-0159.
\end{acknowledgments}

\section*{Appendix}

We recall some basic facts of representation theory needed for the results in this paper and then show how to obtain the double coset sums in Eq.~\ref{eq:double_coset_sums}. Given any finite group $G$ and a complex vector space $V$, a linear map $\Phi:G\rightarrow \text{End}(V)$ which takes the group identity to the identity endomorphism is called a representation of $G$. Often, the space $V$ is called the representation. If there exists a subspace $W$ such that the map $\Phi$ is taken to the subspace for every element $G$, $W$ is called a sub-representation or an invariant subspace of $V$. The orthogonal complement of $W$ in $V$ will also be an invariant space. For any invariant space, the space itself and the trivial subspace consisting of the zero vector is always an invariant space. If the only invariant spaces of $V$ are the trivial space and itself, then $V$ is called an irreducible representation. Every finite group has a finite set of irreducible representations (sometimes called irreps) associated with it. Any representation can be decomposed into irreducible representations, where there may be many copies of a given irreducible representation in it. Invariant spaces of a group can be related to eigenspaes of an operator that is symmetric with respect to that group. More precisely, suppose that a matrix $A$ commutes with a representation of a group $G$ i.e., $U(g)A=AU(g)$ for all $g\in G$. Then the invariant spaces of $G$ lie inside eigenspaces of $A$. Suppose that $U$ has the following block diagonal decomposition
\[
U(g)=\bigoplus_\lambda I_{m_\lambda}\otimes \lambda(g)\,,
\]
where $m_\lambda$ is the multiplicity space of the irrep $\lambda$ i.e., the number of times $\lambda$ appears in $U$. Then $A$ has the decomposition given by
\[
A=\bigoplus_\lambda A_\lambda \otimes I_{d_\lambda} \,,
\]
where $A_\lambda$ is a matrix inside the multiplicity space and $d_\lambda$ is the dimension of $\lambda$. Note that if, in particular, $m_\lambda=1$, then $A_\lambda$ is one-dimensional and therefore is an eigenvalue of $A$. Even if $m_\lambda\neq 1$ but are small, we only need to diagonalize $A_\lambda$ for all $\lambda$ to determine the eigenvalues of $A$.

Now, we give the details of the calculations used to produce Eq.~\ref{eq:double_coset_sums}. The case for the trivial representation is simple. We explain here how to obtain the coset sums for the standard representation (denoted $b$) and the representation we denoted as $c$. We assume familiarity with the representation theory of the symmetric group. The double coset sum for $C_0$ is also trivial to obtain since $\chi_x(C_0)$ (where $x$ represents $b$ or $c$) is the multiplicity of the trivial $G_0$ representation in $x$. This, by Frobenius reciprocity, is the multiplicity of the representation $x$ in the induced representation, which is $1$ for both $b$ and $c$. So, now if we evaluate the double coset sum $\chi_x(C_1)$, then because both $b$ and $c$ are non-trivial irreducible representations, $\chi_x(C_2)=-1-\chi_x(C_1)$. Thus, we only need to evaluate one double coset sum $\chi_x(C_1)$ for $b$ and $c$. In order to do this, note that the double coset sum is actually a multiple of a coset sum i.e., $\chi_x(C_1)=2(N-2)\chi_x((13)G_0)$. Now in order to evaluate $\chi_x((13)G_0)$, find the trivial $G_0$ states in the induced representation. First, we represent the states in the induced representation as $\ket{\{i,j\}}$ (and there are ${N \choose 2}$ of these and they form an orthonormal basis of the induced representation). One can see that there are three trivial $G_0$ states (this follows from Frobenius reciprocity and the fact that this induced representation is multiplicity free). The first one is also a trivial $S_N$ state and is $\sum_{i,j:i\neq j}\ket{\{i,j\}}$ (denote this by $|t\rangle$). The second one comes from the standard $S_N$ representation (i.e., representation $b$). Note that the states in the representation $b$ are given by
\be
\ket{s_i}=\sum_{j:j\neq i}\ket{\{i,j\}} - \frac{2}{N}\ket{t}\,.
\ee
Therefore, the (unnormalized) trivial $G_0$ state in $b$ can be seen to be $\ket{s_1}+\ket{s_2}$ (denote this by $\ket{u}$). Now, we can find $\chi_b(C_1)$ as
\be
\chi_b(C_1)=2(N-2)\frac{\bra{u}(13)\ket{u}}{\langle u |u\rangle}\,.
\ee
This turns out to be $N-4$. Therefore, $\chi_b(C_2)=-(N-3)$. Now, in $c$, the trivial $G_0$ state turns out to be (denoted $\ket{v}$).
\be
\ket{v}=\ket{\{1,2\}} - \frac{1}{N-2}\ket{u} - \frac{1}{{N\choose 2}}\ket{t}\,.
\ee
The double coset sum $\chi_c(C_1)$ can be calculated as follows.
\be
\chi_c(C_1)=2(N-2)\frac{\bra{v}(13)\ket{v}}{\langle v |v\rangle}\,.
\ee
This turns out to be $-2$. This means that $\chi_c(C_2)=1$. One can also double check by calculating $\chi_b(C_2)$ and $\chi_c(C_2)$ independently.
\be
\chi_b(C_2)=\frac{\bra{u}(13)(24)\ket{u}}{\langle u |u\rangle}\,, \chi_c(C_2)=\frac{\bra{v}(13)(24)\ket{v}}{\langle v |v\rangle}\,.
\ee

\end{document}